\documentstyle [12pt]{article}
\newcommand{\id}{\mbox{{\it {\bf 1}}}}
\let\uml=\" \let\acut=\'  
  \def\phi{\varphi}
\def\pa{\partial}
 \def\a{\alpha} 
    
\def\l{\lambda} 
 \def\tr{\mathop{\hbox{\rm tr}}\nolimits}

\def\one#1{#1^{\raise5pt\hbox{$\scriptstyle\!\!\!\!1$}}\,{}}
\def\two#1{#1^{\raise5pt\hbox{$\scriptstyle\!\!\!\!2$}}\,{}}
\def\three#1{#1^{\raise5pt\hbox{$\scriptstyle\!\!\!\!3$}}\,{}}

 \def\comment#1{}
 
\def\endproof{\hfill\rule{2mm}{2mm}}
\def\beq{\begin{equation}} \def\eeq{\end{equation}}
\def\be{\begin{displaymath}} \def\ee{\end{displaymath}}
\def\bea{\begin{eqnarray}} \def\eea{\end{eqnarray}}
\def\beas{\begin{eqnarray*}} \def\eeas{\end{eqnarray*}}
\def\bds{\begin{description}} \def\eds{\end{description}}
\def\bmat{\left(\begin{array}} \def\emat{\end{array}\right)}

\def\Ref#1{(\ref{#1})}
\def\?{(?)\marginpar{|?}}

\renewcommand{\theequation}{\thesection.\arabic{equation}}
\newcounter{subequation}[equation]
\makeatletter
 
\expandafter\let\expandafter\reset@font\csname reset@font\endcsname
 
\def\subeqnarray{\arraycolsep1pt
    \def\@eqnnum\stepcounter##1{\stepcounter{subequation}%
        {\reset@font\rm(\theequation\alph{subequation})}}
\jot5mm     \eqnarray}

\makeatother\newcommand{\newsection}[1]{\vspace{10mm}
\pagebreak[3]\addtocounter{section}{1}\setcounter{equation}{0}
\setcounter{subsection}{0}\setcounter{footnote}{0}
 
\begin{flushleft}{\Large\bf \thesection. #1}
\end{flushleft}\nopagebreak\medskip\nopagebreak}

\newfont{\bbd}{msbm10 scaled\magstep1} \def\C{\hbox{\bbd C}}

\newfont{\frak}{eufm10 scaled\magstep1}

\newcounter{remctr}

\newcommand{\tfrac}[2]{{\textstyle\frac{#1}{#2}}}
\begin{document}
\begin{flushright}
\sf OCTOBER 1996
\end{flushright}
\begin{center} \LARGE\bf
Separation of variables for the ${\cal D}_n$ type periodic
Toda lattice
\end{center}
\vskip 1cm
\begin{center}
Vadim B.~Kuznetsov
\footnote{On leave from: Department of Mathematical and Computational
Physics, Institute of Physics, St.~Petersburg University, St.~Petersburg
198904, Russian Federation. E-mail: {\tt vadim@amsta. leeds.ac.uk}} \\
\vskip 0.7cm
Department of Applied Mathematical Studies,\\  
University of Leeds, Leeds LS2 9JT, UK
\end{center}
\vskip 2.5cm
\begin{center}
\bf Abstract
\end{center}
We prove separation of variables for the most general (${\cal D}_n$ type)
periodic Toda lattice with $2\times 2$ Lax matrix. 
It is achieved by finding proper 
normalisation for the corresponding Baker-Akhiezer function.
Separation of variables for all other periodic Toda lattices 
associated with infinite series of root systems follows by
taking appropriate limits. 
\vskip 2cm
\begin{flushleft}
{\tt solv-int/9701009}
\end{flushleft}
\newpage
\newsection{Introduction}
\setcounter{equation}{0}
Bogoyavlensky \cite{Bo} introduced periodic Toda lattices corresponding
to the root systems of affine algebras. In this case the integrable 
potentials in the Hamiltonian 
\beq
H=\sum_{j=1}^n\frac{p_j^2}{2}+V(q)\,,\qquad \{p_j,q_k\}=\delta_{jk}\,,
\eeq
for the loop algebras ${\cal A}_n^{(1)}$, ${\cal B}_n^{(1)}$, ${\cal C}_n^{(1)}$, 
and ${\cal D}_n^{(1)}$ have the form
\bea
V_{{\cal A}_n^{(1)}}&=&V_{{\cal A}_n}+\exp(q_n-q_1)\,,\nonumber\\
V_{{\cal B}_n^{(1)}}&=&V_{{\cal A}_n}+\exp(q_n)+\exp(-q_1-q_2)\,,\nonumber\\
V_{{\cal C}_n^{(1)}}&=&V_{{\cal A}_n}+\exp(2q_n)+\exp(-2q_1)\,,\nonumber\\
V_{{\cal D}_n^{(1)}}&=&V_{{\cal A}_n}+\exp(q_{n-1}+q_n)+\exp(-q_1-q_2)\,,\nonumber
\eea
where 
$$
V_{{\cal A}_n}=\sum_{j=1}^{n-1}\exp(q_j-q_{j+1})\,.
$$
{}For the twisted loop algebras the integrable potentials are as follows \cite{RS}:
\bea
V_{{\cal A}_{2n}^{(2)}}&=&V_{{\cal A}_n}+\exp(q_n)+\exp(-2q_1)\,,\nonumber\\
V_{{\cal A}_{2n+1}^{(2)}}&=&V_{{\cal A}_n}+\exp(-q_1-q_2)+\exp(2q_n)\,,\nonumber\\
V_{{\cal D}_{n+1}^{(2)}}&=&V_{{\cal A}_n}+\exp(q_n)+\exp(-q_1)\,.\nonumber
\eea

Inozemtsev \cite{I1} found a generic (${\cal D}_n$ type) periodic
Toda lattice with 4 more parameters ($A,\,B,\,C,\,D$) in the potential,
\bea
V(q)&=&V_{{\cal A}_n}+\exp(-q_1-q_2)+\exp(q_{n-1}+q_n)\nonumber\\
&&+\frac{A}{\sinh^2\frac{q_1}{2}}+\frac{B}{\sinh^2q_1}
+\frac{C}{\sinh^2\frac{q_n}{2}}+\frac{D}{\sinh^2q_n}\,,
\label{hhh}\eea
which includes all the above potentials as limiting cases. He gave $2n\times 2n$
Lax representation and proved Liouville integrability for this system. 

Sklyanin \cite{S2} found $2\times 2$ Lax representations for all cases 
(including ${\cal BC}_n$) except 
${\cal B}_n^{(1)}$, ${\cal D}_n^{(1)}$, ${\cal A}_{2n+1}^{(2)}$, and (\ref{hhh}),
introducing
reflection equation which also provided quantisation of those systems. 
The $2\times 2$ Lax matrices ($L$-operators) for the rest 3 cases 
and for Inozemtsev's extension (\ref{hhh}) were found in \cite{K1,K2,K3}. 
See also \cite{K4} where Inozemtsev's case was interpreted as the ${\cal A}_n$ type
open Toda lattice interacting with two Lagrange tops (one on each end of the 
lattice). 

Periodic Toda lattice (of the ${\cal A}_n^{(1)}$ type) was separated 
in \cite{FM}. In \cite{S3} it was treated within the $R$-matrix method
which allowed separation of its quantum counterpart. Partial results
on separation of variables for other Toda lattices were scattered 
in several places \cite{K2,K3,K4}, essentially repeating the basic technique
of \cite{S3} for the case of reflection equation algebra introduced in \cite{S2}. 
As for a detailed algebro-geometrical treatment of many of these Toda lattices 
we refer to \cite{avm}.

In the present paper we prove separation of variables for the 
generic potential (\ref{hhh}) with the $2\times 2$ Lax matrix $L(u)$. 
It is achieved by finding proper normalisation for the 
corresponding Baker-Akhiezer (BA) function $f(u)$
$$
L(u)\;f(u)=v\;f(u)\,,\qquad (\; f(u)=(f_1(u),f_2(u))^t \;)\,. 
$$
We recall that (usually) the separation variables are obtained 
as poles of the BA function (cf. review \cite{S1}). The standard
normalisation $f_1(u)=1$ (or $f_2(u)=1$) which was valid, for instance,
for the ${\cal A}_n^{(1)}$ case \cite{S3} does not work here, giving
too many poles which are not in involution with respect to the Poisson
bracket. The reason is extra symmetries of the Lax matrix. 
To obey the symmetry and reduce the number of poles to 
the number of degrees of freedom, one has to find
a specific normalisation $\vec\a(u)=(\a_1(u),\a_2(u))$ of the BA vector:
$$
\a_1(u)\;f_1(u)+\a_2(u)\;f_2(u)=1\,.
$$

Structure of the paper is following. In Section 2 we give an overview
of the method of separation of variables and apply it then, in the Section 3,
to the integrable system in question. In Section 4 there are 
some concluding remarks. 

\newsection{The method}
\setcounter{equation}{0}
The method of separation of variables plays an important role 
in studying Liouville integrable systems. 
\vskip 0.3cm

\noindent
{\bf Definition 1.} {\it A Liouville integrable system
possesses a Lax matrix if there is a matrix $L(u)$ dependent
on a ``spectral parameter'' $u\in \C$ such that its characteristic 
polynomial obeys two conditions
\bea
&&(i) \quad\mbox{{\it Poisson involutivity:}}\nonumber\\
&&\quad\quad \{\det(L(u)-v\cdot\id),\det(L(\tilde u)-\tilde v\cdot\id)\}=0\,,
\quad\forall u,\tilde u,v,\tilde v\in\C; \nonumber\\
&&(ii)\quad \det(L(u)-v\cdot\id) \quad \mbox{{\it generates all integrals
of motion}}\;\;H_i\,.
\nonumber\eea}

\noindent
{\bf Definition 2.} {\it By separation of variables (SoV) in the classical mechanics
we call an existence 
of a canonical transformation $M:(x,p)\mapsto (u,v)$,
$M:H_i(x,p)\mapsto H_i(u,v)$ such that $H_i(u,v)$ are
in the separated form:
$$
\Phi(u_i,v_i;H_1,\ldots,H_n)\equiv \det(L(u_i)-v_i\cdot\id)=0\,,
\quad i=1,\ldots,n\,.
$$}

\noindent
The above definition corresponds precisely to the standard definition
of SoV in the Hamilton-Jacobi equation \cite{Ar}. 

We would like to notice here that we have connected our definition
of SoV to Lax representation and to associated spectral 
curve of the Lax matrix $L(u)$, so it might be not unique (if exists)
in the case when a chosen integrable system has, for instance,
two or more inequivalent Lax representations.

One of the main questions in the theory is:
How to constructively define those new {\it separation variables}
$(u_j,v_j)$ sitting on the spectral curve of an $L$-matrix
for a given integrable system?

For a very long time a great deal of attention has been 
given to so-called coordinate separation of variables
or to separation in the configuration space
(see, for instance, \cite{Ka,S4,me1,me2,me3,S1} and references therein).
In this case the separation variables $u_j$ are functions 
of $x_i$'s only:
\beq
u_j=u_j(x_1,\ldots,x_n)\,.
\label{6}\eeq
Such kinds of integrable systems admitting a coordinate 
(spatial) separation of variables were studied in detail,
although in the same time it was understood that far not
every Liouville integrable system can be separated through a transition
(\ref{6}) to new ``coordinates'' $u_i$. The class of admissable
transformations should be enlarged for a generic integrable system
upto a general canonical transformation
\beq
u_j=u_j(x_1,\ldots,x_n,p_1,\ldots,p_n)\,,\qquad 
v_j=v_j(x_1,\ldots,x_n,p_1,\ldots,p_n)\,.
\label{7}\eeq
The very existence of SoV
according to the above definition is still unproved in general,
to author's knowledge; although there are powerful methods
which have been applied to many families of integrable systems
(see recent review \cite{S1}) showing that separability is one of the 
most important features of integrability, and that hopefully
latter always implies former. 
The method of SoV in its modern formulation can be found in 
\cite{S1}. See also the works \cite{KS1,KS2,KS3,KS4,KNS}.
Here we describe very briefly its main steps.

The first difficulty is: {\it How to find separation variables
$u_j$?} There is a general answer to this question, 
which has been inspired 
by the whole experience of the inverse scattering method, and it is a very
simple one:
\vskip 0.3cm

\noindent
{\it \underline{Answer:} They [$u_j$] are poles of the Baker-Akhiezer 
function which is properly normalized.}
\vskip 0.3cm

\noindent
There is, however, a slight further problem of choosing the right normalisation
for the BA function; the problem which was not completely 
solved by powerful and successful method of inverse scattering.
So, a general theory connecting the symmetry of the Lax matrix to 
proper normalisation vector of the BA function is still incomplete.
But, supposing that one somehow knows the right normalisation,
then one could proceed further and put the above general 
recipe into the formulas (cf. \cite{S1}).

The linear problem for the BA function $f(u)$ is of the form
\beq
L(u)\;f(u)=v(u)\;f(u)\,,\qquad (\;\det(L(u)-v\cdot\id)=0\;)\,.
\label{5.5}\eeq
The normalisation $\vec\a(u)$ of the eigenvectors $f(u)$ has to be fixed
\beq
\sum_{i=1}^N\a_i(u)\;f_i(u)=1\,,\qquad (\;f(u)\equiv
(f_1(u),\ldots,f_N(u))^t\;)\,.
\label{5.6}\eeq
Let $L(u)$ be a meromorphic function in $u$ then $f(u)$ is also meromorphic
in $u$. Let us look at its [$f(u)$'s] poles $u_j$:
$$
f_i^{(j)}=\mbox{{\rm res}}_{u=u_j}f_i(u)\,.
$$
Then from (\ref{5.5})--(\ref{5.6}) we have 
\beq
\left\{\matrix{
L(u_j)\;f^{(j)}=v_j\;f^{(j)}\,,\qquad v_j\equiv v(u_j)\,,\cr {}\cr
\sum_{i=1}^N\a_i(u_j)\;f_i^{(j)}=0\,.}\right.
\label{5.7}\eeq
Equations (\ref{5.7}) are $N+1$ linear homogeneous
equations for the separation variables
$u=u_j$ and $v=v_j$ which are bounded by definition to the spectral curve
(cf. \Ref{5.5}). These equations have to be compartible. 
The system (\ref{5.7}) is equivalent to the 
condition:
\beq
\mbox{{\rm rank}}\pmatrix{\vec\a(u)\cr L(u)-v\cdot\id}=N-1
\label{5.9}\eeq
where $\vec\a$ is thought of as a row-vector.
Finally, the condition (\ref{5.9}) can be rewritten as the 
following vector equation:
\beq
\vec \a\cdot (L(u)-v\cdot\id)^\wedge=0\,,
\label{5.11}\eeq
where wedge denotes the classical adjoint matrix (matrix of cofactors). 
\vskip 0.3cm

\noindent
{\bf Proposition 1} {\it 
Excluding $v$, one can derive from equations (\ref{5.11}) the equation for $u$ 
in the form
\beq
B(u)=\det\pmatrix{\vec\a\cr
\vec\a \cdot L(u)\cr
\vdots\cr
\vec\a \cdot L^{N-1}(u)}=0\,.
\label{5.12}\eeq}
\vskip 0.3cm

\noindent
{\bf Proof}
When $u=u_j$ we have the equations (cf. (\ref{5.7}))
\beq
L(u)\;f=v(u)\;f\,,\qquad {\vec\alpha}\,f=0\,.
\label{17}\eeq
Hence
\be
{\vec\alpha}\;L^k\,f=0\,,\qquad k=0,1,2,\ldots\,.
\ee
Then (\ref{5.12}) follows because $f$ is a non-zero vector.
\endproof
\vskip 0.3cm

\noindent
Also, from equations (\ref{5.11}) we can get formulas for $v$ in the form
$$
v=A(u)
$$
with $A(u)$ being some rational functions of the entries of $L(u)$ (cf. \cite{KNS}). 

What is left is just to verify (somehow) the canonical brackets between
the whole set of separation variables, namely: between zeros $u_j$ of $B(u)$ 
and their conjugated variables $v_j\equiv v(u_j)=A(u_j)$. 
To do this final calculation we need information about Poisson brackets between
entries of the Lax matrix $L(u)$ which is usually provided by 
corresponding $r$-matrix (standard or dynamical). 

In order to perform
a SoV, say, in a strong sense, one has to try also to get an explicit 
representation for the corresponding generating function $F(u|x)$
of the separating canonical transform $M$ from the set $(x_j,p_j)$
to the set $(u_j,v_j)$. Actually, to find the generating function
$F(u|x)$ one has to solve the system of non-linear equations
of the form
$$
\vec\a(u_j)\;(L(u_j)+\tfrac{\pa F}{\pa u_j}\cdot\id)^\wedge\;
{}|_{p_k=\frac{\pa F}{\pa x_k}}=0\,,\qquad j=1,\ldots,n\,.
$$
In the quantum case the function $F(u|x)$ has a quantum counterpart:
the kernel ${\cal M}_\hbar(u|x)$ of the separating integral transform
$M_\hbar$, so that
$$
{\cal M}_\hbar(u|x)\sim \exp(\tfrac{i}{\hbar}F(u|x))\,,\qquad \hbar\rightarrow0\,.
$$
For some integrable systems such 
special functions of many variables ($F$ and ${\cal M}_\hbar$) 
can be obtained in very explicit terms (cf. \cite{S1,KS1,KS2,KS3}). 
Remark here that 
the other generating function has been traditionally associated to 
constructions of the method of separation of variables in the 
Hamilton-Jacobi equation, namely: the action function $S(H|u)$ given in
terms of separation variables, $u_j$, and integrals of motion, $H_j$. 
Our choice 
of arguments of the generating function is justified by the quantum case
where $F(y|x)$ has a direct quantum analog, while the action function
$S(H|u)$ does not have such a nice quantum counterpart at all. 

Very often the above prescription of SoV should be read 
``in the opposite direction'' (because one does not usually know 
the separating normalisation in advance). 
Sometimes, regardless of choosing
the vector ${\vec \alpha}$, the Baker-Akhiezer
function $f(u)$ has just needed number of poles
in involution. Sometimes, and this is very important, $f(u)$
has too many poles and they are \underline{not} mutually in involution,
showing that there are some constraints between them. In the latter case,
one should find proper (and quite unique)
normalisation vector ${\vec \alpha}(u)$
so that to fix all extra poles of $f(u)$ being \underline{constants}.  
The prescription then makes us to search for a way to resolve
possible constraints on poles of the Baker-Akhiezer function 
by using the freedom of choosing its normalisation. 
In this paper we show that it is the case in the ${\cal D}_n$ type periodic
Toda lattice and give the right normalisation for corresponding
$f(u)$, thereby producing a SoV for this system which was not solved
before by this method. 

If we make a similarity transformation for the $L$-matrix
\be
\widetilde{L}(u)=V(u)\;L(u)\;V^{-1}(u)
\ee
with a non-degenerate matrix $V(u)$ then the linear problem
\beq
L(u)\;f(u)=v(u)\;f(u)\,,\qquad
\vec\alpha\cdot f=1
\label{19}\eeq
turns into 
\be
\widetilde{L}(u)\;\widetilde{f}(u)=v(u)
\;\widetilde{f}(u)\,,\qquad
{\vec\alpha}_0\cdot\widetilde{f}=1
\ee
where 
\beq
\widetilde{f}(u)=V(u)\;f(u)\,,\qquad {\vec\alpha}(u)=
{\vec\alpha}_0(u)\;V(u)\,.
\label{more}\eeq
This shows that the freedom of choosing the normalisation vector $\vec\alpha$
is equivalent to the freedom of making similarity transformations
to the initial Lax representation. 

Let us put $N=2$, so that we assume from now on that we have 
a $2\times 2$ Lax representation for our integrable system. 
In this case the equations of SoV (\ref{5.9}) have the form
\be
\mbox{{\rm rank}}\pmatrix{\a_1(u)&\a_2(u)\cr 
                          L_{11}(u)-v&L_{12}(u)\cr
                          L_{21}(u)&L_{22}(u)-v}=1\,.
\ee
{}From which we conclude that 
$$
\left\{\matrix{\a_1\,L_{12}=\a_2\,(L_{11}-v)\cr {} \cr
\a_1\,(L_{22}-v)=\a_2\,L_{21}}
\right.\Leftrightarrow \left\{\matrix{B(u)=\a_1^2L_{12}-\a_1\a_2(L_{11}-L_{22})
-\a_2^2L_{21}=0\cr {} \cr v=A(u)=L_{11}-\tfrac{\a_1}{\a_2}\,L_{12}=
L_{22}-\tfrac{\a_2}{\a_1}\,L_{21}}\right.\,.
$$
Suppose we have found a non-degenerate matrix $V(u)$ such that
the Lax matrix $\widetilde{L}(u)=V(u)L(u)V^{-1}(u)$ ends up in 
SoV with the standard normalisation vector $\vec\a_0=(1,0)$. That would
imply the separability for the matrix $L(u)$ with the normalisation vector
$\vec\a$ (cf. (\ref{more}))
$$
\vec\a=\vec\a_0\cdot V=(V_{11}(u),V_{12}(u))\,.
$$%
\newsection{The separation}
\setcounter{equation}{0}
Let us remind first the construction of the $2\times 2$ Lax matrix for the 
${\cal D}_n$ type periodic Toda lattice with four extra parameters (Inozemtsev's
case) \cite{K1,K2,K3,K4}. 

Given the rational classical $4\times 4$ $r$-matrix of the form
\be
r(u)=\frac{\kappa}{u}\pmatrix{1&0&0&0\cr 0&0&1&0\cr
0&1&0&0\cr 0&0&0&1}\,,
\ee
one considers two algebras: the Sklyanin quadratic algebra
\be
(S)\; \qquad 
\{L^{(1)}(u),L^{(2)}(v)\}=[\,r(u-v),L^{(1)}(u)\,L^{(2)}(v)\,]\,,
\qquad\qquad\ee
and the reflection equation algebra
\be
(RE)\qquad 
\{L^{(1)}(u),L^{(2)}(v)\}=[\,r(u-v),L^{(1)}(u)\,L^{(2)}(v)\,]
\qquad\qquad\ee
\be \qquad\qquad\qquad\qquad\qquad 
+L^{(1)}(u)\,r(u+v)\,L^{(2)}(v)-L^{(2)}(v)\,r(u+v)\,L^{(1)}(u)\,.
\ee
These two algebras appeared in the quantum inverse scattering method.
Their representaions play an important role in the classification
and studies of classical integrable systems (see, for
instance, \cite{FT,RS,S2,S1} and references in there).
Here the sup-indices $(1)$ and $(2)$ mean standard tensoring of the $2\times 2$
matrix $L(u)$ with the $2\times 2$ unit matrix $\id$: 
$L^{(2)}(u)=\id\otimes L(u)$, $L^{(1)}(u)=L(u)\otimes\id$.

The following $2\times 2$ $L$-operators 
\be
L_1(u)=\left(\matrix{
u^2x_1+u\,[\, i(x_1^2-1)\,p_1+c_1x_1+c_2\,]+c_1c_2\cr
u\,(\,u^2+(x_1^2-1)\,p_1^2-2ip_1\,(c_1x_1+c_2)-c_1^2\,)}\right.
\qquad\qquad\qquad\qquad
\ee
\be
\qquad\qquad\qquad\qquad\qquad\qquad\left.\matrix{u\,(x_1^2-1)\cr
u^2x_1-u\,[\,i(x_1^2-1)\,p_1+c_1x_1+c_2\,]+c_1c_2}\right)
\ee
and
\be
L_2(u)=\left(\matrix{
-u^2x_2+u\,[\,i(x_2^2-1)\,p_2+c_3x_2+c_4\,]-c_3c_4\cr
u\,(x_2^2-1)}\right.\qquad\qquad\qquad\qquad
\ee
\be
\qquad\qquad\qquad\qquad\qquad\qquad\left.\matrix{
u\,(\,u^2+(x_2^2-1)\,p_2^2-2ip_2\,(c_3x_2+c_4)-c_3^2\,)\cr
-u^2x_2-u\,[\,i(x_2^2-1)\,p_2+c_3x_2+c_4\,]-c_3c_4}\right)
\ee
satisfy the $(RE)$ algebra with $\kappa=i$. 
Here the $(x_j,p_j)$ are canonical Darboux variables, i.e. the Poisson
brackets are $\{p_j,x_k\}=\delta_{jk}$.  
These $L$-operators were found in \cite{K1,K2} (see also \cite{K3,K4}).
They generate the ${\cal D}_n$ type periodic Toda lattice having four additional
(singular) potential terms with the parameters $c_1,\,c_2,\,c_3,\,c_4$. 
Namely, consider the following Lax matrix 
\beq
T(u)=L_3(u)\cdot\ldots\cdot L_n(u)\;\cdot\; L_1(u)\;
\cdot \;L_n^{-1}(-u)\cdot\ldots\cdot L_3^{-1}(-u)\;\cdot\; L_2(u)\,,
\label{hhhh}\eeq
where the $L$-operators $L_3,\ldots,L_n$ satisfy the $(S)$
algebra with $\kappa=i$ and have the form:
\be
L_k(u)=\pmatrix{0&-x_k^{-1}\cr x_k&u+ip_kx_k}\,,\qquad 
k=3,\ldots,n\,.
\ee
The constructed Lax matrix 
describes an integrable system with the following Hamiltonian:
\bea
H_1&=&\sum_{i=3}^n \;(x_ip_i)^2 + p_1^2\,(x_1^2-1)+p_2^2\,(x_2^2-1)
-2\sum_{i=3}^{n-1}\frac{x_i}{x_{i+1}}\nonumber\\
&&+2\,\frac{x_2}{x_3}+2x_1x_N-2ip_1\,(c_1x_1+c_2)-2ip_2\,(c_3x_2+c_4)\,.
\label{o1}\eea
This Hamiltonian turns into the one for Inozemtsev's Toda lattice (cf. (\ref{hhh}))
under the following change of variables:
$x_1=\cosh\,q_1,\;\;x_2=\cosh\,q_2,\;\;x_j=\exp(q_j),\;j=3,\ldots,n$,
and obvious gauge-type canonical transformation for two particles
(with the variables $(x_1,p_1)$ and $(x_2,p_2)$) 
to get rid of terms linear in $p_1,p_2$ in (\ref{o1}). 

Our problem is to separate variables in this system and 
restore the Lax matrix $T(u)$ (\ref{hhhh}) 
in terms of (new) separation variables. This is perfomed in the following
three Propositions.

Spectral curve has the following form:
\be
\mbox{det}\,(T(u)-v\cdot\id)
\ee
\be
=v^2-v\left[
(-1)^nu^{2n+2}+(-1)^nH_1\,u^{2n}+H_2\,u^{2n-2}+\ldots+H_n\,u^2-
2c_1c_2c_3c_4\right]\ee
\beq
+\prod_{i=k}^4(u^2-c_k^2)=0\,.
\label{dett}\eeq
\vskip 0.3cm

\noindent
{\bf Proposition 2} {\it 
Let
\beq
V(u)=\pmatrix{1-x_2&u+c_3-ip_2\,(1-x_2)\cr 0&\frac{1}{1-x_2}}\,.
\eeq
Then it is easy to verify that $V(u)$ obeys the (S) algebra with $\kappa=i$
and, moreover, it sends the matrix $L_2(u)$ into 
the triangular form:
\be
\widetilde{L}_2(u)\equiv V(-u)\cdot L_2(u)\cdot V^{-1}(u)=\pmatrix{
(u-c_3)(u+c_4)&0\cr -u\;\frac{1+x_2}{1-x_2}&(u+c_3)(u-c_4)}\,.
\ee}
\vskip 0.3cm

\noindent
{\bf Proof}
It is a simple and straightforward algebraic calculation. The second
part of the statement is crucial for the following procedure of
separation of variables
and is absolutely non-trivial since we apply {\it almost} similarity
transformation to the boundary matrix $L_2(u)$ to put it into the
triangular form (notice the changed sign of the spectral
parameter $u$).  
\endproof
\vskip 0.3cm

\noindent
{\bf Proposition 3} {\it 
Consider the representation of the (RE) algebra 
of the following form:
\be
\widetilde{T}(u)=V(u)\cdot
L_3(u)\cdot\ldots\cdot L_n(u)\;\cdot\; L_1(u)\;
\cdot \;L_n^{-1}(-u)\cdot\ldots\cdot L_3^{-1}(-u)\cdot V^{-1}(-u)
\ee
\beq
=\pmatrix{\widetilde{A}(u)&\widetilde{B}(u)\cr
\widetilde{C}(u)&\widetilde{D}(u)}\,.
\eeq
Then the matrix $\widehat{T}(u)$ which is similar to the $T(u)$ 
can be represented as follows:
\beq
\widehat{T}(u)\equiv V(u)\cdot T(u)\cdot V^{-1}(u)=
\widetilde{T}(u)\cdot \widetilde{L}_2(u)\,. 
\label{decomp}\eeq
Hence
\be
\mbox{{\rm tr}}\,T(u)=
(u-c_3)(u+c_4)\;\widetilde{A}(u)+(u+c_3)(u-c_4)\;\widetilde{D}(u)
-u\;\frac{1+x_2}{1-x_2}\;\widetilde{B}(u)\,,
\ee
with
\be
\mbox{{\rm det}}\,\widetilde{T}(u)  =(u^2-c_1^2)(u^2-c_2^2)\,,\qquad
\mbox{{\rm det}}\,\widetilde{L}_2(u)=(u^2-c_3^2)(u^2-c_4^2)\,,\ee
\be
\mbox{{\rm det}}\,T(u)=\prod_{k=1}^4(u^2-c_k^2)\,.
\ee
If we choose $n$ zeros $u_k$ of the polynomial
$\widetilde{B}(u)$ as $n$ separation variables:
\beq
\widetilde{B}(\pm u_k)=0\,,\qquad \lambda_k^\pm=\widetilde{D}(\pm u_k)\,,
\qquad k=1,\ldots,n\,, 
\label{BBB}\eeq
then they satisfy the relations
\be
\{u_j,u_k\}=0\,,
\ee
\be
\{u_k,\lambda_k^\pm\}=\pm i\;\lambda_k^\pm\,,
\ee
\be
\lambda_k^+\;\lambda_k^-=(u_k^2-c_1^2)\,(u_k^2-c_2^2)\,,
\ee
\be
\{\lambda_j^\pm,\lambda_k^\pm\}=
\{\lambda_j^\pm,\lambda_k^\mp\}=
\{\lambda_j^\pm,u_k\}=0\,,\qquad j\neq k\,.
\ee
Moreover, from their definition it follows that they
satisfy the equalities ($k=1,\ldots$, $n$)
\beq
\mbox{{\rm tr}}\,T(u_k)=(u_k-c_3)(u_k+c_4)\;\lambda_k^-+
(u_k+c_3)(u_k-c_4)\;\lambda_k^+
\label{BBB1}\eeq
(the separation equations).}
\vskip 0.3cm

\noindent
{\bf Proof} 
The matrix $\widetilde{T}(u)$ satisfies the involution
\be
\widetilde{T}(-u)=\left[\mbox{det}\,\widetilde{T}(u)\right]\cdot
\, \widetilde{T}^{-1}(u)=
\sigma_2\;\widetilde{T}^t(u)\;\sigma_2
\ee
or, in component-wise form,
\be
\widetilde{A}(-u)=\widetilde{D}(u)\,,\qquad
\widetilde{B}(-u)=-\widetilde{B}(u)\,,\qquad
\widetilde{C}(-u)=-\widetilde{C}(u)\,.
\ee
Moreover, its polynomial in $u$ entries have the degrees
\be
\mbox{deg}\,\widetilde{T}(u)=\pmatrix{2n&2n+1\cr 2n-1&2n}\,.
\ee
The matrix $\widetilde{T}(u)$ obeys the $(RE)$ algebra of Poisson
brackets according to the Proposition 2 from \cite{S2} because $L_j(u)$, 
$j=3,\ldots,n$, and $V(u)$ obey the $(S)$ algebra brackets. Using our 
Proposition 2 we establish the decomposition (\ref{decomp}) for the matrix
$\widehat{T}(u)$ which is similar to the Lax matrix $T(u)$. The rest of the 
formulas are obvious. The polynomial $\widetilde{B}(u)$ has exactly $n$
non-trivial zeros $u_k$, $k=1,\ldots,n$ (doubled by the obvious 
$\pm$-symmetry). The related $\l_k^\pm$ variables are defined 
according to (\ref{BBB}). These new variables $u_k,\;\l_k^\pm$ are bounded 
to the equalities (\ref{BBB1}) by their definition. The calculation of
all the Poisson brackets between the separation variables $u_k,\;\l_k^\pm$
is a standard procedure nowadays which was originally invented in \cite{S3}.
Let us recall, for instance, how one calculates the brackets between
$u_k$ and $\l_k^+$. From the $(RE)$ algebra for $\widetilde{T}(u)$ 
we have
$$
-i\{\widetilde{B}(u),\widetilde{D}(v)\}=
\frac{\widetilde{D}(u)\widetilde{B}(v)-\widetilde{D}(v)\widetilde{B}(u)}
{u-v}+
\frac{\widetilde{D}(-u)\widetilde{B}(v)+\widetilde{D}(v)\widetilde{B}(u)}
{u+v}\,.
$$
Combining it with the equation
$$
0=\{\widetilde{B}(u_k),\widetilde{D}(v)\}=\{\widetilde{B}(u),\widetilde{D}(v)\}\;{}
|_{u=u_k}+\widetilde{B}'(u_k)\{u_k,\widetilde{D}(v)\}
$$
we obtain
$$
\{u_k,\l_k^+\}=\tfrac{-i}{\widetilde{B}'(u_k)}\;(\tfrac{\l_k^+}{u_k-v}
+\tfrac{\l_k^-}{u_k+v})\;\widetilde{B}(v)\;{}|_{v=u_k}=i\,\l_k^+\,.
$$
\endproof
\vskip 0.3cm

\noindent
{\bf Proposition 4} {\it 
The interpolation problem to restore the matrix
$\widetilde{T}(u)$ in terms of new (separation) variables
$u_k,\lambda_k^\pm$ has the following solution:
\be
\widetilde{B}(u)=(-1)^nu\prod_{k=1}^n(u^2-u_k^2)\,,
\ee
\be
\widetilde{D}(u)=(-1)^nc_1c_2\prod_{k=1}^n
\frac{u^2-u_k^2}{u_k^2}
+\sum_{k=1}^n\left[\frac{u(u+u_k)}{2u_k^2}\,\lambda_k^++
\frac{u(u-u_k)}{2u_k^2}\,\lambda_k^-\right]\prod_{j\neq k}
\frac{u^2-u_j^2}{u_k^2-u_j^2}\,,
\ee
\be
\widetilde{A}(u)=\widetilde{D}(-u)\,,\qquad \widetilde{C}(u)=
\frac{\widetilde{A}(u)\widetilde{D}(u)-(u^2-c_1^2)(u^2-c_2^2)}
{\widetilde{B}(u)}\,.
\ee}
\vskip 0.3cm

\noindent
{\bf Proof} 
The formula for $\widetilde{B}(u)$ is obvious. The polynomial
$\widetilde{D}(u)$ of degree $2n$ is restored in terms of the 
separation variables by interpolation with $2n+1$ data 
of the form
$$
\widetilde{D}(\pm u_k)=\l_k^\pm\,, \qquad \widetilde{D}(0)=c_1c_2\,.
$$
\endproof
\vskip 0.3cm

\noindent
Now we can derive, in principle, the formulas connecting old
and new variables. For instance, noticing that $\widetilde{D}(u)$
has the asymptotics
\be
\widetilde{D}(u)=\frac{(-1)^n}{1-x_2}\;u^{2n}+\ldots\,,
\qquad u\rightarrow\infty\,,
\ee
we find that 
\be
\frac{1}{1-x_2}=\frac{c_1c_2}{\prod_{k=1}^n u_k^2}
+(-1)^n\sum_{k=1}^n\frac{\lambda_k^++\lambda_k^-}
{2u_k^2\prod_{j\neq k}(u_k^2-u_j^2)}\,.
\ee
We can express some other combinations of initial variables 
in terms of new (separation) variables,
comparing the coefficients of entries of $\widetilde{T}(u)$ 
in both representations. Considering the $\mbox{tr}\,T(u)$,
we could as well get the expressions for the integrals of motion
$H_1,\ldots,H_n$ in terms of the separation variables.
\vskip 0.3cm

\noindent
{\bf Corollary 1} {\it The separating normalisation vector
for the ${\cal D}_n$ type periodic Toda lattice 
with the Hamiltonian (\ref{o1}) and with the Lax matrix (\ref{hhhh})
has the form
$$
\vec\a=(\,1-x_2,\;u+c_3-ip_2\,(1-x_2)\,)\,.
$$
The separation variables $u_k$ and $v_k^\pm\equiv (u_k\pm c_3)(u_k\mp c_4)\,
\l_k^\pm$, $k=1,\ldots,n$,
are sitting on the spectral curve (\ref{dett}) of the Lax matrix $T(u)$ 
(\ref{hhhh})
$$
(v_k^\pm)^2-v_k^\pm\,\tr\,T(u_k)+\det\,T(u_k)=0\,,
$$
i.e.
$$
v_k^++v_k^-=\tr\,T(u_k)\,,\qquad v_k^+\,v_k^-=\det\,T(u_k)\,.
$$
They have the following Poisson brackets
$$
\{u_k,v_k^\pm\}=\pm i\;v_k^\pm\,.
$$}
\vskip 0.2cm

\noindent
{\bf Remark 1} {\it The (obvious) alternative choice of the separating
normalisation vector follows if we put the matrix $L_1(u)$ instead
of the matrix $L_2(u)$ (cf. Proposition 2) into the triangular form. 
This would correspond to interchanging two edge particles in the lattice.} 
\vskip 0.4cm

It would be interesting to (explicitly) construct the generating
function $F(u|x)$ of this separating canonical transform. 

If we introduce the canonically conjugate variables $\pi_j$
$$
\{\pi_j,u_k\}=\delta_{jk}
$$
then we can put
$$
v_k^\pm=[\det\,T(u_k)]^{\tfrac12}\;\exp(\mp i\,\pi_k)
$$
and get the separation equations in the form
$$
2\,[\det\,T(u_k)]^{\tfrac12}\;\cos(\pi_k)=\tr\,T(u_k)\,.
$$
Hence, the action variables $S_k(H_1,\ldots,H_n)$ have the form 
$$
S_k(H_1,\ldots,H_n)=\oint_{\a_k}\mbox{arccos}\,\left(\frac{\tr\,T(u)}
{2[\det\,T(u)]^{\tfrac12}}\right)\,du\,,\qquad k=1,\ldots,n\,,
$$
where $\a_k$ are the $\a$-cycles on the Riemannian surface 
of $\sqrt{\tr^2T(u)-4\det\,T(u)}$. 

One can get the quasiclassical spectrum $H_k(N_1,\ldots,N_n)$ of the 
integrals of motion $H_1,\ldots,H_n$ (cf. \cite{KKK}) inverting the integrals
(Bohr-Sommerfeld quantisation) 
$$
S_k(H_1,\ldots,H_n)=h\,N_k\,,\qquad k=1,\ldots,n\,,
$$
where $N_k$'s are the quantum numbers, $N_k=1,2,3,\ldots$.
Obtaining of true discrete spectrum of the integrals of quantum ${\cal D}_n$ type
periodic Toda lattice is the problem of quantum separation of variables. 

\newpage
\newsection{Concluding remarks}
We refer reader to the review \cite{S1} (cf. also the work \cite{KNS})
where it was illustrated
that the simplest choice of the normalisation vector ${\vec\alpha}$,
when one of the components of the Baker-Akhiezer function $f(u)$ (for instance
the first one) is equal to 1, i.e. when 
\beq
{\vec\alpha}=(1,0,\ldots,0)\,,
\label{18}\eeq
provides a SoV for many integrable systems of the ${\cal A}_n$ type. If a chosen
integrable system can not be separated with this simplest normalisation, 
and this usually means that its $L$-matrix has some extra symmetries/involutions
(i.e. is of the ${\cal BC}_n$ or ${\cal D}_n$ type or obeys elliptic $r$-matrix),
then the main problem is to find proper ${\vec\alpha}$.
For the time being there is no theory to give a general prescription for 
finding right normalisation vector ${\vec\alpha}$ in those cases. 
Although one practical rule can be suggested. Usually, if one looks
at the poles of the Baker-Akhiezer function with the simplest normalisation
(\ref{18}), one finds that there are too many poles and they
do not respect the symmetry presenting in the problem. 
Then the rule is the following: take an ansatz for ${\vec\alpha}(u)$
with some dependence on $u$ and with some indeterminates in it, derive 
equations for those indeterminates demanding that (a) $f(u)$ with such
a normalisation has the right number of moving poles respecting involutions 
of the spectral curve and (b) all extra poles are equal to constants. Then solve
the equations ... . 

In this paper we applied this approach to the ${\cal D}_n$ type 
periodic Toda lattice with four additional singular terms in the potential.
This system is not separated with the simplest choice of the 
normalisation vector ${\vec\alpha}$ (\ref{18}), so we have derived the right
normalisation ${\vec\alpha}$ producing the SoV. For some of the root 
systems the separating normalisation vector
is a constant vector (cf. the ${\cal BC}_n$ case in \cite{K2,K3,K4}). 
{}For the generic ${\cal D}_n$ 
case the separating ${\vec\alpha}(u)$ depends on
the spectral parameter $u$ and on the phase variables, so it is dynamical.
We think that it is an important feature of this kind of problems
(the ones with extra involutions),
that the separating choice of ${\vec\alpha}$ is not quite arbitrary,
as it was for some of the ${\cal A}_n$ type of systems, but is quite unique
and dynamical. 

The specific situation with the ${\cal D}_n$ type periodic Toda lattice, i.e.
that the right ${\vec\alpha}$ is $u$-dependent and dynamical, is surely
connected with the fact that we have the dynamical boundary $L_{1,2}$-matrices
in construction of the corresponding Lax matrix $T(u)$ (\ref{hhhh}) 
for this case. 

\section*{Acknowledgments}
The main part of this work was done while the author was visiting Centre de
recherches math\' ema\-ti\-ques (CRM), Universit\' e de Montr\' eal. 
The author wishes to acknowledge the support of the CRM-ISM and EPSRC. 

\newpage
\bibliographystyle{plain} 

\begin{thebibliography}{99}
\bibitem{avm} M.~Adler and P.~van~Moerbeke, {\it Completely integrable systems, 
Euclidean Lie algebras, and curves}, Adv. Math. {\bf 38} (1980), 267--317; 
{\it Linearization of Hamiltonian systems, Jacobi varieties and representation 
theory}, ibid. 318--379.
\bibitem{Ar} V.~I.~Arnol'd, Mathematical methods of classical mechanics,
Springer, New-York Heidelberg Berlin, 1974.
\bibitem{Bo} O.~I.~Bogoyavlensky, {\it On perturbation of the periodic
Toda lattices}, Commun. Math. Phys. {\bf 51} (1976), 201--209.
\bibitem{me3} J.~C.~Eilbeck, V.~Z.~Enol'skii, V.~B.~Kuznetsov and A.~V.~Tsiganov, 
{\it Linear $r$-matrix algebra for classical separable systems,} 
J.\ Phys.\ A:\ Math.\ Gen.\  {\bf 27} (1994), 567--578.
\bibitem{FT} L.~D.~Faddeev and L.~A.~Takhtajan,
Hamiltonian methods in the theory of solitons,
Springer, Berlin, 1987.
\bibitem{FM} H.~Flaschka and D.~McLaughlin, {\it Canonically
conjugate variables for the Korteweg-de Vries equation and the 
Toda lattice with periodic boundary conditions}, Prog. Theor. Phys.
{\bf 55} (1976), 438--456.
\bibitem{I1} V.~I.~Inozemtsev, {\it The finite Toda lattices}, 
Commun. Math. Phys. {\bf 121} (1989), 629--638.
\bibitem{Ka} E.~G.~Kalnins, Separation of variables for Riemannian
spaces of constant curvature, Pitman Monographs and Surveys in Pure
and Applied Mathematics {\bf 28}, Longman Scientific and Technical,
Essex, England, 1986.
\bibitem{KKK} I.~V.~Komarov and V.~B.~Kuznetsov,
{\it Quasiclassical quantisation of the Kovalevskaya top},
Theoret. and Math. Phys. {\bf 73} (1987), 1255--1263.
\bibitem{K1} V.~B.~Kuznetsov, Applications of the inverse scattering 
method to 2-dimensional classical and quantum integrable dynamical systems,
Ph.D. Thesis, Leningrad University (1989), 118pp.
\bibitem{K2} V.~B.~Kuznetsov, {\it Generalized polyspheroidal periodic functions and 
the quantum inverse scattering method}, J. Math. Phys. {\bf 31} (1990), 1167--1174.
\bibitem{me1} V.~B.~Kuznetsov, {\it Quadrics on real Riemannian spaces of constant
curvature: separation of variables and connection with Gaudin magnet,}
J.\ Math.\ Phys.\ {\bf 33} (1992), 3240--3254.
\bibitem{me2} V.~B.~Kuznetsov, {\it Equivalence of two graphical calculi,}
J.\ Phys.\ A:\ Math.\ Gen.\  {\bf 25} (1992), 6005--6026.
\bibitem{K4} V.~B.~Kuznetsov, M.~F.~J\o rgensen and P.~L.~Christiansen,
{\it New boundary conditions for integrable lattices}, J. Phys. A: Math. Gen. 
{\bf 28} (1995), 4639--4654.
\bibitem{KNS} V.~B.~Kuznetsov, F.~W.~Nijhoff and E.~K.~Sklyanin, {\it Separation 
of variables for the Ruijsenaars system}, CNLS-Leeds preprint (1997), 
{\tt solv-int/9701004}.
\bibitem{KS1} V.~B.~Kuznetsov and E.~K.~Sklyanin, {\it Separation of
variables in $A_2$ type Jack polynomials},
RIMS Kokyuroku {\bf 919} (1995), 27--34.
\bibitem{KS2} V.~B.~Kuznetsov and E.~K.~Sklyanin, {\it Separation of
variables for the $A_2$ Ruijsenaars model and a new integral
representation for the $A_2$ Macdonald polynomials},
J. Phys. A: Math. Gen. {\bf 29} (1996), 2779--2804.
\bibitem{KS3} V.~B.~Kuznetsov and E.~K.~Sklyanin, {\it Separation of
variables and integral relations for special functions} (1996). Submitted.
\bibitem{KS4} V.~B.~Kuznetsov and E.~K.~Sklyanin, {\it Factorisation 
of Macdonald polynomials}, In: Proceedings of the Second Workshop
on Symmetries and Integrability of Difference Equations (SIDEII), 
July 1996, Canterbury, UK, to appear.
\bibitem{K3} V.~B.~Kuznetsov and A.~V.~Tsiganov, {\it Infinite series of Lie algebras
and boundary conditions for integrable systems}, J. Sov. Math. {\bf 59} (1992),
1085--1092.
\bibitem{RS} A.~G.~Reyman and M.~A.~Semenov-Tian-Shansky,
Group theoretical methods in the theory of finite dimensional
integrable systems, in {\it Dynamical systems VII}, Encyclopaedia of Math. Sci.,
Springer {\bf 16} (1994).
\bibitem{S3} E.~K.~Sklyanin, {\it The quantum Toda chain}, in Non-linear
equations in classical and quantum field theory, Lecture Notes in Phys.
{\bf 226}, ed. N.~Sanchez (Springer, 1985), p.196.
\bibitem{S2} E.~K.~Sklyanin, {\it Boundary conditions for integrable 
quantum systems}, J. Phys. A: Math. Gen. {\bf 21} (1988), 2375--2389.
\bibitem{S4} E.~K.~Sklyanin, {\it Separation of variables in the Gaudin model},
J. Sov. Math. {\bf 47} (1989), 2473--2488.  
\bibitem{S1} E.~K.~Sklyanin, {\it Separation of variables. New trends,}
Progr.\ Theor.\ Phys.\ Suppl.\ {\bf 118} (1995), 35--60.
\end{thebibliography}

\end{document}